\documentstyle[11pt,a4,cite,epsf]{article}

\newcommand{\be}{\begin{equation}}
\newcommand{\ee}{\end{equation}}
\newcommand{\bea}{\begin{eqnarray}}
\newcommand{\eea}{\end{eqnarray}}
\newcommand{\noi}{\noindent}
\newcommand{\nn}{\nonumber}

\begin{document}

\begin{titlepage}
\begin{flushleft} {\Large\sl Revised version}
\end{flushleft}
\begin{flushright} CPT-95/P3202\\ CERN-TH/95-141
\end{flushright}
\vspace{2cm}
\begin{center} {\Large \bf Two--Loop Electroweak Corrections to the Muon
g--2\,:}

{\Large \bf a new class of Hadronic Contributions }\\[1.5cm]  {\large {\bf
Santiago Peris}$^{a,b}$\footnote{On leave from Grup de F\'{\i}sica Te\'orica
and IFAE, Universitat Aut\'onoma de Barcelona, Barcelona, Spain. E-mail: PERIS
@ SURYA11.CERN.CH}\footnote{Work partially supported by research project
CICYT-AEN93-0474.},  {\bf Michel Perrottet}$^a$ and {\bf Eduardo de
Rafael}$^a$}\\[1cm]
${}^a$  Centre  de Physique Th\'eorique\\
       CNRS-Luminy, Case 907\\
    F-13288 Marseille Cedex 9, France\\[0.5cm] and\\[0.5cm]
$^b$ Theoretical Physics Division, CERN,\\
     CH-1211 Geneva 23, Switzerland.\\
\end{center}
\vspace*{1.5cm}
\begin{abstract}

We discuss, within the framework of the Standard Model, the calculation
of the
two-loop electroweak contributions to the anomalous magnetic moment of
the muon
involving triangle fermionic loops of  leptons and  quarks. Because
of the large ratios of masses involved, these contributions are rather
large. The result we obtain differs from a previous estimate
reported in the literature. The discrepancy originates in the cancellation of
anomalies in
$SU(3)_c\times SU(2)_L
\times U(1)_Y$, a cancellation that requires the consideration of
both leptons {\it and} quarks within each generation and that had been
previously overlooked.

\end{abstract}
\vfill
\begin{flushleft} CERN-TH/95-141\\
\end{flushleft}

\begin{flushleft} hep-ph/9505405\\
\end{flushleft}

\begin{flushleft} May 1995\\
\end{flushleft} \end{titlepage}

{\bf 1.} There
is a forthcoming experiment at the Brookhaven National Laboratory which plans
to measure the anomalous magnetic moment of the muon with an expected
uncertainty of $\pm 40\times 10^{-11}$. This
will correspond to an improvement by a factor of {\it twenty} with respect to
the latest result obtained from the experiment performed at CERN, which gave
\cite{BFP90}:

\be\label{eq:CERN} a_{\mu}\equiv \frac{1}{2}(g_{\mu}-2) =
11659230(85)\times10^{-10}.
\ee
\noi The muon $g-2$ project at BNL \cite{Hug92} has renewed the interest on the
part of some theorists to improve the accuracy of the corresponding prediction
in the Standard Model. Here we want to concentrate on the electroweak
higher-order corrections. The history and the present status of the calculation
of the
dominant electromagnetic contributions to
$a_{\mu}$ can be found in the series of review articles \cite{PLdeR72},
\cite{CNPdeR77}, and \cite{KM90}. A recent phenomenological re-evaluation of
the hadronic vacuum polarization effect on
$a_{\mu}$ has been made in ref.\cite{EJ95}. The hadronic light--by--light
scattering effect on
$a_{\mu}$ has also been recently reconsidered, within the framework of
low--energy QCD, in refs.\cite{deR94}, \cite{HKS95} and \cite{BPP95}.

The one-loop contributions to $a_{\mu}$ due to the electroweak interactions of
the Standard Model were calculated quite a long time ago \cite{BGL72}. The
relevant Feynman diagrams are shown in Fig.1. The corresponding result is

\be \label{eq:a1loop} a_{\mu}^{\rm Weak}=\frac{G_{\rm F}}{\sqrt
2}\frac{m_\mu^2}{8\pi^2}\Biggl\{
\frac{10}{3}
+\frac{4}{3}(v_{\mu}^2-5\, a_{\mu}^2) + {\cal
O}\biggl(\frac{m_{\mu}^2}{M_Z^2}
\log\frac{M_Z^2}{m_{\mu}^2}\biggr) + 2\int_{0}^{1}
dx\frac{x^2(2-x)}{x^2+\frac{M_{\rm H}^2}{m_{\mu}^2}(1-x)} \Biggr\} \,,
\ee
\noi where $v_{\mu}$ and $a_{\mu}$ are the vector and axial--vector couplings
of the
$Z$ to the muon. In general, for a fermion $f$\,,

\be \label{eq:couplings} v_{f}=I_f^{(3)}-2Q_{f}\sin^2\theta_W\,, \qquad
a_{f}=I_f^{(3)}\,.
\ee
\noi The contribution from the Higgs, given in terms of a parametric integral
in  eq. (\ref{eq:a1loop}), decouples in  the infinite mass limit. Numerically,
with the Higgs contribution neglected,

\be\label{eq:number1} a_{\mu}^{\rm Weak}=195\times 10^{-11}\,.
\ee

The leading  two--loop electroweak contributions to $a_{\mu}$ have been
discussed in ref.\cite{KKSS92}. These authors have selected all the possible
sources of logarithmically enhanced terms which appear because of the large
ratios of masses involved, and which result in contributions to
$a_{\mu}$  of order

\be {\cal O}\biggl(\frac{G_{\rm F}}{\sqrt 2}\frac{m_\mu^2}{8\pi^2}\,
\frac{\alpha}{\pi}\, \log\frac{M^2}{m^2}\biggr)\,,
\ee
\noi with $M\gg m$, where typically $M$ is the $Z$ mass and $m$ a
fermion mass, like for instance the muon. Apart from these,
other possible logarithms involving the Higgs mass
like $\log M_H/M_Z$ or $\log M_H/m_t$ might also appear but, unless
$M_H\gg m_t$, they are not so large and are disregarded.
The authors of ref. \cite{KKSS92} thus find the following overall
correction $\Delta
a_{\mu}^{\rm Weak}$ to the one--loop result in eq. (\ref{eq:number1}):

\be
\label{eq:russian-estimate}
\Delta a_{\mu}^{\rm Weak}\simeq -42\times 10^{-11}\,;
\ee
\noi i.e. a rather large negative correction, of the same size as the
expected experimental uncertainty \cite{Hug92}.

{\bf 2.} We shall be concerned with a specific class of the two-loop
electroweak contributions: those induced by virtual fermionic triangle loops,
represented by the Feynman diagrams in Fig. 2. The authors of ref.
\cite{KKSS92}
have only considered the subclass of these  contributions where the fermion in
the triangle loop is a {\it lepton}. In the `t Hooft-Feynman gauge and
keeping only the asymptotic contributions from the large ratios of
masses involved, the results they find are:

\bea
\label{eq:electron}
\Delta a_{\mu}^{\rm Weak}\vert_{e}& \simeq & -\frac{G_{\rm F}}{\sqrt
2}\frac{m_\mu^2}{8\pi^2}\,
\frac{\alpha}{\pi}\biggl[ 3\log\frac{M_Z^2}{m_{\mu}^2}+\frac{5}{2}\biggr]  =
-11.7\times 10^{-11}\,;\\ \label{eq:muon}
\Delta a_{\mu}^{\rm Weak}\vert_{\mu}& \simeq & -\frac{G_{\rm F}}{\sqrt
2}\frac{m_\mu^2}{8\pi^2}\,
\frac{\alpha}{\pi}\biggl[3\log\frac{M_Z^2}{m_{\mu}^2}-\frac{8}{9}\pi^2
+\frac{11}{6}\biggr]  =  -9.11\times 10^{-11}\,;\\ \label{eq:tau}
\Delta a_{\mu}^{\rm Weak}\vert_{\tau}& \simeq & -\frac{G_{\rm F}}{\sqrt
2}\frac{m_\mu^2}{8\pi^2}\,
\frac{\alpha}{\pi}\biggl[3\log\frac{M_Z^2}{m_{\tau}^2}-6\biggr]  =
-4.77\times 10^{-11}\,.
\eea

One should realize that the results of eqs. (\ref{eq:electron}-\ref{eq:tau})
are
actually gauge dependent. These results, as they stand, stem from the
$g_{\mu \nu}$ part of the $Z$ propagator in the diagrams of Fig. 2.
\footnote{The
associated diagrams with a would-be Nambu-Goldstone have an extra
suppression due to the mass of the fermion going around the triangle,
$m_f$, that goes like $m_f^2/M_Z^2$.}. In the unitary gauge, for instance,
the $k^{\mu} k^{\nu}$ piece in the $Z$ propagator yields an extra
contribution to eqs. (\ref{eq:electron}-\ref{eq:tau}) that has a part that
is {\it common} to these three equations. This extra common
contribution is actually {\it divergent} and originates in the anomaly that
results when one multiplies the triangle by $k^{\mu}$. It is only when one
sums over a complete generation that the anomaly vanishes and the result is
finite and gauge invariant.\footnote{The gauge invariance of the Standard
Model is a fact only after the cancellation of anomalies is effected.} As a
consequence, strictly speaking, only the contribution of a full
generation may be considered physically meaningful, but not that of a
single fermion.

The Feynman diagrams in Fig.2 where the fermion in the triangle loop is a
{\it quark} correspond to a new class of
{\it hadronic contributions} to the muon
anomaly. We shall call them the hadronic
$Z$--$\gamma$--$\gamma$ contributions.
The fact that the fermionic triangle
subdiagram in Fig. 2 has an Adler, Bell--Jackiw
$VVA$ anomaly, which in the Standard Model cancels when all the fermions of
the same generation are included, implies the vanishing of the
whole triangle diagram in the limit of exact mass degeneracy within each
generation. We therefore expect important cancellations within
each generation,
which questions the overall estimate in eq. (\ref{eq:russian-estimate})
quoted from  ref.\cite{KKSS92}.

We propose to investigate this problem first within the
framework of effective quantum
field theories where the underlying physics can be easily understood. We have
also done an exact calculation of the contribution to
$a_{\mu}$ from the diagrams in Fig. 2, and checked that, in the appropriate
cases, the various asymptotic limits reproduce the simple effective field
theory expressions. We have reproduced in particular the results in
eqs. (\ref{eq:electron}), (\ref{eq:muon}), and (\ref{eq:tau}) in the `t
Hooft-Feynman gauge.  We shall then discuss the new numerical results.


\vspace{7 mm} {\bf 3.} Let us first consider the limit where the $Z$--mass is
much larger than the masses of any of the other particles involved, which is
certainly the case for the first and second generations, and keep only the
leading $\log M_Z$ contributions. When the
$Z$--field in the Standard Model is integrated out, there appear new local
four--fermion couplings induced by the tree--level exchange of the underlying
$Z$ propagator. These four--fermion couplings lead to two--loop diagrams like
the ones shown in Fig.3, which yield contributions to the muon
$g-2$ that are logarithmically divergent. This divergence is to be
interpreted as cut off by $M_Z$. If furthermore the fermion masses in the
triangle loop are neglected
with respect to the muon mass, the result from each fermionic contribution is
the same, up to a factor $Q_f^2a_f$ proportional to the square of the electric
charge
$Q_f$ of the fermion times its axial coupling $a_f$ defined in
eq. (\ref{eq:couplings}). For quarks, there is an extra factor
of {\it three} from the number of colours. The overall result
we find for the muon
$g-2$ from the first generation, in the limit where $m_e =0$, and in
the chiral
limit where $m_u =m_d =0$, is then:

\be \label{eq:1stgen}
\Delta a_{\mu}^{\rm Weak}\vert_{e,d,u}\simeq \frac{G_{\rm F}}{\sqrt
2}\frac{m_\mu^2}{8\pi^2}\,
\frac{\alpha}{\pi}\sum_f Q_f^2a_f\,
\left(6\log\frac{M_Z^2}{m_{\mu}^2}+{{\cal A}\over \epsilon} + {\cal B}\right)\,
=0\, ,
\ee
\noi
where the $1/\epsilon$ term encodes the ultraviolet divergences in
dimensional regularization and the $m_f$-independent constants ${\cal A},
{\cal B}$ are gauge dependent. They
vanish in the `t Hooft-Feynman gauge but not in the unitary gauge.
Each fermion contribution is separately gauge dependent and it
is only the sum over the full generation that is physically meaningful.
The fact that $\sum_f Q_f^2a_f=0$ is of course a property due to the anomaly
cancellation in the $SU(3)_c\times SU(2)_L\times U(1)_Y$ gauge theory.
We shall
later come back to a more elaborate discussion of the contribution from the
first generation, which takes into account the effect of hadronic mass
scales in the light quark sector of QCD.

\vspace{7 mm}
{\bf 4.} The evaluation of the leading contribution from the
second generation of fermions is a little more delicate. We can still consider
the effective field theory where the $Z$--field has been integrated out, but
now the fermion masses in the triangle loop cannot be neglected with respect to
the external muon mass. The contribution from the strange quark, in particular,
requires a special discussion depending on whether or not one is willing to
consider the chiral
$SU(3)$ limit where $m_s=0$, and on whether or not one wants to discuss
spontaneous chiral symmetry breaking effects and the effects due to the QCD
$U(1)_{A}$ anomaly. To a first approximation, we shall also consider the chiral
limit for the strange quark and, as in the case of the first-generation
estimate, we shall neglect for the time being the effect of hadronic mass
scales. We then find :

\bea
\Delta a_{\mu}^{\rm Weak}\vert_{\mu ,s,c} & \simeq & \frac{G_{\rm F}}{\sqrt
2}\frac{m_\mu^2}{8\pi^2}\,\frac{\alpha}{\pi}\biggl\{(-\frac{1}{2})
6\log\frac{M_Z^2}{m_{\mu}^2} + 3 \frac{1}{9}
(-\frac{1}{2}) 6\log\frac{M_Z^2}{m_{\mu}^2} +
3\frac{4}{9}(\frac{1}{2}) 6\log\frac{M_Z^2}{m_c^2} \biggr\} \nn \\
\label{eq:2ndgen}
 & = & -\frac{G_{\rm F}}{\sqrt 2}\frac{m_\mu^2}{8\pi^2}\,\frac{\alpha}{\pi}\,
4\log\frac{m_c^2}{m_{\mu}^2}\simeq -5.4\times 10^{-11}\,,
\eea
\noi where the numerical result is for $m_c\!=\!1.3\,$GeV. As expected
on first principles, the $Z$--mass does not appear in the final result.

\vspace{7 mm}
{\bf 5.} The evaluation of the leading contribution from the
third generation brings in an interesting issue, related to the fact that the
top quark is heavier than the
$Z$. Within the framework of effective field theories, we now have to consider
the case where the top field is integrated out first, corresponding to
the limit
$m_{t}\gg M_Z$. In this limit, the top quark in the $Z$--$\gamma$--$\gamma$
vertex decouples. The corresponding effective local $Z$--$\gamma$--$\gamma$
coupling induced by the top triangle loop goes as $1/m_t^2$ and
induces a contribution to the muon $g-2$ at the one--loop level
via the Feynman diagrams shown in Fig. 4. If we work in the limit $m_t\to
\infty$, this contribution vanishes. In this limit and in the unitary gauge
(where there are no would-be Nambu-Goldstone bosons) the integration of the
top leaves no trace behind in the form of new local effective operators
relevant to the $g_{\mu}-2$.\footnote{There are of course other operators
relevant to other processes.\cite{Santa}}

In the effective Lagrangian without top the $g_{\mu \nu}$ part of the $Z$
propagator
yields for $\tau$ and $b$ in the loop a finite contribution completely
analogous to that found in the previous case for the first two
generations. The $k^{\mu} k^{\nu}$ part, on the other hand, yields the
anomaly when multiplied by the triangle and this contributes a
logarithmically divergent quantity that is to be interpreted as cut off by
the top mass. Gathering all the pieces we obtain:

\bea
\label{eq:top}
\Delta a_{\mu}^{\rm Weak}\vert_{\tau,b,t} & \simeq &  -\frac{G_{\rm F}}{\sqrt
2}\frac{m_\mu^2}{8\pi^2}\,\frac{\alpha}{\pi}\biggl\{
3\log\frac{M_Z^2}{m_{\tau}^2}+\log\frac{M_Z^2}{m_b^2} - \sum_{f=\tau,b}
Q_f^2 a_f\ 4\log\frac{m_t^2}{M_Z^2}\biggr\} \nn \\
 & = & - 9\times 10^{-11}\,,
\eea
\noi where we have used  $m_{\tau}\!=\!1.78\,$GeV, $m_b\!=\!4.3\,$GeV, and
$m_t\!=\!170\,$GeV for the numerical estimate. The last term in this
equation comes from
the $k^{\mu} k^{\nu}$ part of the $Z$
propagator. This last term with the sum over $\tau$ and $b$ would have been
the contribution from the top quark had we done the calculation with the
full theory, i.e. without integrating out the top.

Altogether, treating the light $u$, $d$, and $s$ quarks in the chiral
limit, and
neglecting the effect of hadronic mass scales, we find that in the Standard
Model the leading contribution to the muon $g-2$ from the full set of
fermionic triangle graphs in Fig. 2 represents a correction of $\sim -7\%$ to
the dominant one--loop electroweak contribution.

Of course the limit $m_t\to \infty$ is quite far away from the real
situation, and in the next section we will present exact expressions that
will give corrections to this extreme limit.

\vspace{7 mm}
{\bf 6.} The effective field theory framework is not very useful when
non-leading effects in the large ratios of masses are also required. In that
case it is simpler to do a direct calculation. The result corresponding to
the diagrams in Fig. 2, in the unitary gauge, for a fixed fermion $f$ in the
triangle loop and without approximations has the form

\be
\label{eq:fermion}
\Delta a_{\mu}^{\rm Weak}\vert_f = -\frac{G_{\rm F}}{\sqrt
2}\frac{m_\mu^2}{8\pi^2}\,\frac{\alpha}{\pi}Q_f^2 a_f\,\biggl\{ {\cal F} [
\frac{m_f^2}{m_{\mu}^2},\frac{M_Z^2}{m_{\mu}^2}]
 - \frac{m_f^2}{M_Z^2}{\cal G} [
\frac{m_f^2}{m_{\mu}^2},\frac{M_Z^2}{m_{\mu}^2}] \biggr\}\,,
\ee
\noi  where we have separated the contributions
generated by the
$g_{\mu\nu}$ term in the $Z$ propagator (the ${\cal F}$--function)  from
those generated by the
$k_\mu k_\nu$ term (the ${\cal G}$--function). As already mentioned
earlier, it is the
$k_\mu k_\nu$ term which also generates a divergent piece,
independent of the fermion mass in the loop, and which here has been
subtracted. (It cancels of course, once the sum over the fermions of a
generation is made.) The functions ${\cal F}$ and
${\cal G}$  have rather compact Feynman parametric representations. With $x$
and $y$ the Feynman parameters of the triangle loop, and
$u$,$v$, and $w$ those of the other loop we have:

\bea
{\cal F}[\frac{m_f^2}{m_{\mu}^2},\frac{M_Z^2}{m_{\mu}^2}] & = &
\int_0^1 \!dw\int_0^1 \!dv\int_0^1 \!du\int_0^1\!dx\int_0^{1-x}\!dy
\frac{8u^{2}v}{u^2 v^2 w^2\frac{m_{\mu}^2}{M_Z^2}+
\frac{u(1-v)}{y(1-y)}\frac{m_f^2}{M_Z^2}+(1-u)}\nn \\ \label{eq:exactg}  & &
\times\biggl\{ \frac{2x}{1-y}-3(1+uvw) +
\frac{u^3 v^3 w^3}{u^2 v^2 w^2+
\frac{u(1-v)}{y(1-y)}\frac{m_f^2}{m_{\mu}^2}+(1-u)\frac{M_Z^2}{m_{\mu}^2}}
\biggr\} \,,
\eea
\noi
and

$$ {\cal G}[\frac{m_f^2}{m_{\mu}^2},\frac{M_Z^2}{m_{\mu}^2}]  =
\int_0^1 \!dw\int_0^1 \!dv\int_0^1 \!du\int_0^1\!dx\int_0^{1-x}\!dy
\frac{8u^{2}v}{u^2 v^2 w^2\frac{m_{\mu}^2}{M_Z^2}+
\frac{u(1-v)}{y(1-y)}\frac{m_f^2}{M_Z^2}+(1-u)}
$$

\be \label{eq:exactk}
\times\biggl\{ -\frac{1+3uvw}{y(1-y)} +
\frac{u^3 v^3 w^3}{u^2 v^2 w^2 y (1-y)+
u(1-v)\frac{m_f^2}{m_{\mu}^2}+(1-u)y(1-y)\frac{M_Z^2}{m_{\mu}^2}}
\biggr\} \,.
\ee

The contribution from  the ${\cal G}$--function to the muon $g-2$ is only
sizeable for the $t$ quark. It is in fact this piece which reproduces the
leading
$\log\frac{m_t^2}{M_Z^2}$ discussed earlier. We can now obtain the
non--leading behaviour as well, with the result:

\be
\Delta a_{\mu}^{\rm Weak}[t]\vert_{{\rm eq}.(\ref{eq:fermion})}=
 - \frac{G_{\rm F}}{\sqrt 2}\frac{m_\mu^2}{8\pi^2}\,\frac{\alpha}{\pi}
\biggl\{
\frac{8}{3}\log\frac{m_t^2}{M_Z^2} + \frac{16}{3} -
\frac{2}{9}\frac{M_Z^2}{m_t^2}\log\frac{m_t^2}{M_Z^2} +  {\cal
O}\biggl(\frac{M_Z^2}{m_t^2}\biggr)\biggr\}\,.
\ee

We can also use the expressions above to obtain a more accurate evaluation of
the contributions to $\Delta a_{\mu}^{\rm Weak}$ when the fermion
$f$ in the triangle loop is one of the three leptons\,: $f\!=\!e,
\mu, \tau$. When the fermion
$f$ in the triangle loop is a quark, the expression in eq. (\ref{eq:fermion})
corresponds to the evaluation one obtains in the limit
where the QCD gluonic interactions are neglected. If the quark in the triangle
loop is a {\it heavy quark}\,: $f\!=\!c,b,t$, the QCD gluonic interactions can
be treated perturbatively, and the corrections to the lowest-order estimate
obtained, will be down by a typical factor
$\alpha_{s}(\mu^2)/{\pi}$, where the mass scale $\mu$ runs between $m_f$ and
$M_Z$ \cite{P95}. We can therefore use eq. (\ref{eq:fermion}) to reliably
obtain
a more elaborate estimate of the contribution to the muon
$g-2$ from the third generation of fermions, with the result\,:

\be \label{eq:topimp}
\Delta a_{\mu}^{\rm Weak}[\,\tau,b,t\,]\vert_{{\rm
eq}.(\ref{eq:fermion})}= - 8.2\times 10^{-11}\,,
\ee
\noi to be compared with our approximate estimate in eq. (\ref{eq:top}).

\vspace{7mm} {\bf 7.} When the fermion $f$ in the triangle loop is a
{\it light quark}\,:
$f\!=\!u,d,s$, the result obtained from a straightforward application
of the expression in eq.(\ref{eq:fermion}) with either  current algebra quark
masses or  constituent quark masses can be rather misleading. QCD
perturbation theory  is not justified in this case; and it would be erroneous
to use it since, among other things, it neglects the fact that chiral symmetry
is spontaneously broken. An appropriate way
to discuss this problem is within the combined framework of chiral perturbation
theory ($\chi$PT) and the $1/N_c$ expansion. (For a recent review, where
earlier references can also be found, see e.g. ref.\cite{deR95}.) To lowest
order in the chiral expansion in $U(3)_L\times U(3)_R$, the hadronic
$Z$--$\gamma$--$\gamma$ interaction appears via the one Goldstone meson
exchanges between the effective coupling

\be \label{eq:op2}
{\cal L}^{(2)}=\frac{-{\rm e}}{2\sin\vartheta_{\rm W}\cos\vartheta_{\rm
W}}\,f_{\pi}\,
\partial_{\mu}\biggl(\pi^0 +\frac{1}{\sqrt{3}}\eta_8 -
\frac{1}{\sqrt{6}}\eta_0\biggr)Z^{\mu}\,,
\ee
\noi
induced by the lowest ${\cal O}(p^2)$ chiral effective Lagrangian, and
the effective ${\cal O}(p^4)$ coupling

\be \label{eq:WZ}
{\cal L}_{ABJ} =
\frac{\alpha}{\pi}\frac{N_c}{24f_{\pi}}\biggl(\pi^0+\frac{1}{\sqrt{3}}
\eta_8
+2\sqrt{\frac{2}{3}}\eta_0\biggr)
\epsilon_{\mu\nu\rho\sigma}F^{\mu\nu}F^{\rho\sigma}\,;
\ee
\noi i.e. the term in the Wess--Zumino Lagrangian which reproduces the
Adler, Bell--Jackiw anomaly. The corresponding Feynman diagrams are shown in
Fig.5. The evaluation of the contribution to the muon $g-2$ from these
diagrams, in the unitary gauge and in the chiral limit, leads to the result\,:

\be
\label{eq:chiral}
\Delta a_{\mu}^{\rm Weak}[u,d,s]\vert_{\chi{\rm PT}}=
\frac{G_{\rm F}}{\sqrt 2}\frac{m_\mu^2}{8\pi^2}\,
\frac{\alpha}{\pi}\biggl[\frac{4}{3}\log\frac{M_Z^2}{m_{\mu}^2}+\frac{4}{9}
+{\cal O}\biggl(\frac{m_{\mu}^2}{M_Z^2}\log\frac{M_Z^2}{m_{\mu}^2}\biggr)
\biggr]=5.0\times
10^{-11}\,,
\ee
\noi
where we have also subtracted the divergent piece generated by the
$k_{\mu}k_{\nu}$ term in the $Z$ propagator, which cancels with
the corresponding pieces generated by the $e$, $\mu$, and $c$ fermionic loops.
We wish to emphasize that to lowest non--trivial order in
$\chi$PT, in the chiral limit and in the large
$N_c$ limit of QCD, this is an exact result. When added
together with the contributions to
$\Delta a_{\mu}^{\rm Weak}\vert_{f}$ in eq. (\ref{eq:fermion})  from the
leptons $e$, $\mu$, and the $c$ quark in the triangle loop, it gives as the
result to the contribution to
$\Delta a_{\mu}^{\rm Weak}$ from the first and second generation of leptons
and quarks\,:

\be \label{eq:1st2ndgen}
\Delta a_{\mu}^{\rm Weak}[\,e,d,u; \mu,s,c\,]=-8.7\times
10^{-11}\,,
\ee
\noi
to be compared with our first estimate, the sum from
eqs. (\ref{eq:1stgen}) and (\ref{eq:2ndgen}).


\vspace{7mm} {\bf 8.} We now want to discuss the sources of corrections to the
above calculation and their possible estimate.

One source is the effect of the $\alpha_{s}(\mu^2)/{\pi}$ perturbative
corrections to the lowest-order $c$,
$b$, and $t$ loop calculations which we have already mentioned. They are, in
principle, under control and one does not expect them to drastically change
our results.

The other sources of corrections concern the light-quark sector. Here,
besides the obvious change in the Nambu-Goldstone propagator due to
finite pseudoscalar masses, we
expect chiral loop corrections due to the explicit breaking of the chiral
symmetry, as well
as corrections due to the contributions from higher order terms in the
effective chiral Lagrangian. Chiral-loop corrections are suppressed in the
$1/N_c$ expansion. We expect them to give corrections ${\cal
O}(\frac{m_{\eta}^2}{16\pi^{2}f_{\pi}^2})$, perhaps enhanced by chiral
logarithmic factors. However, the most important effect to next to leading
order in the $1/N_c$ expansion is likely to be the fact that, because of the
$U(1)_A$ anomaly, the singlet $\eta_0$ particle acquires a large mass. The
effect of this mass will be to damp the contribution of the $\eta'$ to the
muon $g-2$ in eq. (\ref{eq:chiral}).

Concerning the effect due to higher order terms in the chiral
expansion, it is possible to make estimates  using
models of the QCD low--energy effective action at large
$N_c$, which have been developed during the last few years (see e.g.
ref.\cite{deR95}.) The simplest version of these models amounts in practice to
giving a constituent mass $M_Q$ to the $u$, $d$, and $s$ quarks, and to
modulate the axial--vector coupling of the constituent quarks with a constant
$g_A$
\cite{BBdeR93}, \cite{PdeR93}. Here, care must be taken, however, on the
way
constituent quark masses and the coupling $g_A$ are introduced. As already
mentioned, the VVA vertex in the triangle loop in Fig. 2 has an anomalous Ward
identity\,: in the chiral limit the VV$\partial$A vertex has a universal form
which is, in particular, at the origin of the effective chiral realization
discussed above. With
$k_{\mu}$ the momentum flowing from the axial coupling, and
$T^{\mu\alpha\beta}$ the full VVA vertex, we can decompose this vertex as
follows\,:

\be
\label{eq:decomp}
T^{\mu\alpha\beta}=\frac{1}{k^2}k^{\mu}k_{\rho}T^{\rho\alpha\beta}+
\biggl(T^{\mu\alpha\beta}-
\frac{1}{k^2}k^{\mu}k_{\rho}T^{\rho\alpha\beta}\equiv
R^{\mu\alpha\beta}\biggr)\,.
\ee
\noi
The first term in the r.h.s., when modulated by the appropriate
$\sum_{f}Q_f^{2}a_f$ factor, with $f\!=\!u,d,s$ and including the $N_c$
colour factor reproduces, in the chiral limit, the expression obtained
in the effective chiral realization that we have already discussed (see
eqs. (\ref{eq:op2}) and (\ref{eq:WZ}).) This is an example of the `t Hooft
anomaly matching condition \cite{`tH80} applied to QCD. The constituent chiral
quark picture should then be applied to the second term only, i.e. the
$R^{\mu\alpha\beta}$ tensor defined as the difference in
eq. (\ref{eq:decomp}). (Notice that the contribution to the
muon $g-2$ from the
$R$--tensor is gauge invariant.) This way, and using the values
$M_Q\simeq (265\pm 10)\,{\rm MeV}$ and  $g_A\simeq 0.6\pm 0.1$, we find
a correction of at most $\simeq 30\%$ to the  result in
eq. (\ref{eq:chiral}). From this we conclude that $50\%$ is a safe estimate of
the size of the expected errors in the
$\chi$PT calculation in eq. (\ref{eq:chiral}).

Our final estimate of the leading (in the sense of eq. (5))
electroweak contributions from the full set of
fermionic triangle loops in the
Standard Model, from eqs. (\ref{eq:topimp}) and
(\ref{eq:1st2ndgen}) is then\,:

\be
\Delta a_{\mu}^{\rm Weak}[\,e,d,u;\, \mu,s,c;\, \tau,b,t\,]=-(16.9\pm
2.5)\times 10^{-11}.
\ee
\noi
This result, when added to the other leading two-loop electroweak
corrections calculated in \cite{KKSS92}, corresponds to an overall estimate

\be
\Delta a_{\mu}^{\rm Weak}\simeq -(36.9\pm 2.5)\times 10^{-11}\,;
\ee
\noi i.e. a negative correction of $\sim 19\%$ to the lowest order electroweak
contribution in eq. (\ref{eq:number1}).

\vspace{0.2 in}

S.P. would like to thank E. Alvarez, L. Alvarez-Gaum\'e and S. Yankielowicz
for conversations. We would like to thank A. Czarnecki for pointing out a
sign error in the contribution of the top in a previous version of the
manuscript.

\vspace{1cm}
\noindent{\it Note Added}

After completion of this work we have become aware of a paper by
A. Czarnecki, B. Krause and W.J. Marciano, hep-ph/9506256, where
a subset of the constant (i.e. non-$\log M/m$ enhanced, see eq. (5))
contributions has been calculated. These contributions originate mainly in
$m_t^2$--like corrections; they are negative and push
the final result in eq. (24) to $-45\times 10^{-11}$.


\vspace{10mm}

\newpage

\noi {\Large \bf Figure Captions}

\vspace{7mm}
\noi Fig.1\,: Feynman diagrams which, in the Standard Model, give the lowest
          order electroweak contribution to the muon $g-2$.
\vspace{3mm}

\noi Fig.2\,: Two--loop electroweak corrections induced by virtual fermionic
          triangle loops.
\vspace{3mm}

\noi Fig.3\,: Two--loop electroweak contributions, with virtual fermionic
          triangle loops, induced by a local four--fermion effective coupling.
\vspace{3mm}

\noi Fig.4\,: One--loop electroweak contribution induced by an effective
          $Z$--$\gamma$--$\gamma$ local coupling.
\vspace{3mm}

\noi Fig.5\,: Lowest order contribution in the effective chiral realization of
          QCD, which leads to a $Z$--$\gamma$--$\gamma$ hadronic contribution
          to the muon $g-2$.


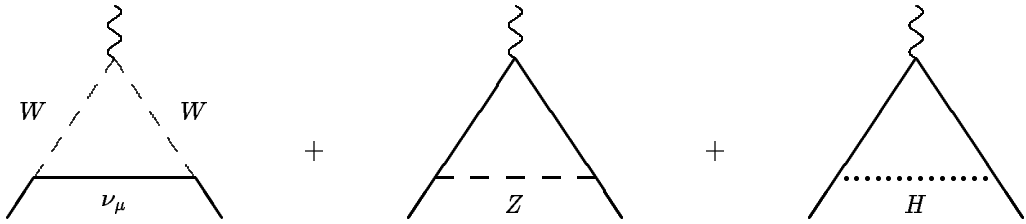
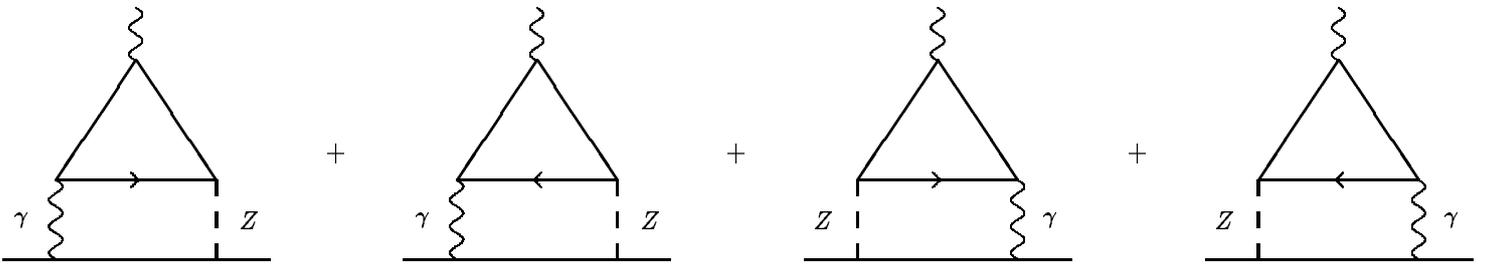
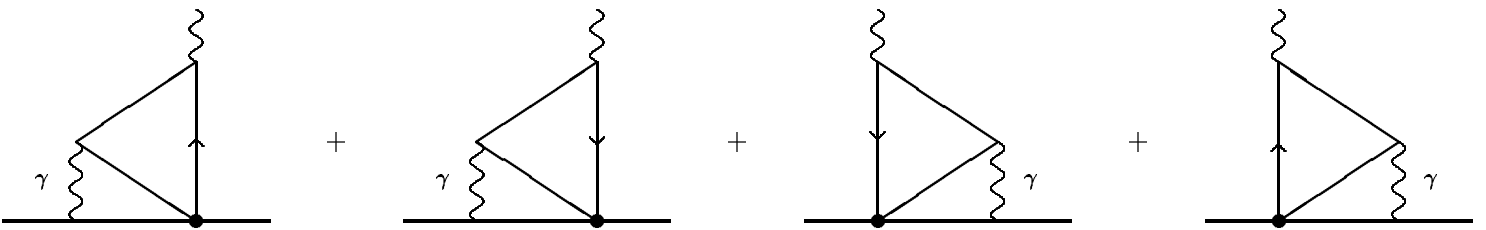
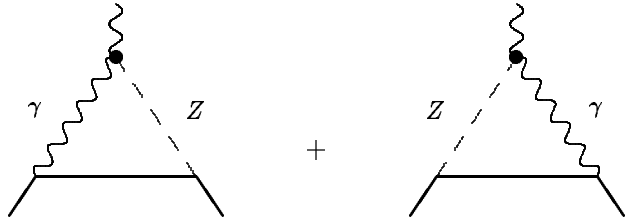
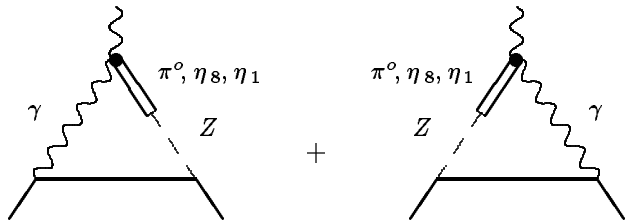
\end{document}